\documentclass[sigconf]{acmart}

\usepackage{booktabs}
\usepackage{tabularx}

\copyrightyear{2026}
\acmYear{2026}
\setcopyright{cc}
\setcctype{by}
\acmConference[ICMI Companion '26]{Companion of the INTERNATIONAL CONFERENCE ON MULTIMODAL INTERACTION}{October 05--09, 2026}{Napoli, Italy}
\acmBooktitle{Companion of the INTERNATIONAL CONFERENCE ON MULTIMODAL INTERACTION (ICMI Companion '26), October 05--09, 2026, Napoli, Italy}
\acmDOI{10.1145/3776591.3837051}
\acmISBN{979-8-4007-2319-3/2026/10}

\begin{document}

\title{Beyond Call and Response: Modelling Reciprocal Coordination in
  Human--AI Vocal Ensembles}

\author{Polina Proutskova}
\orcid{0000-0002-7514-2711}
\affiliation{%
  \institution{Industry Commons Foundation}
  \city{Stockholm}
  \country{Sweden}
}
\email{polina.proutskova@industrycommons.net}
\renewcommand{\shortauthors}{Proutskova}

\begin{abstract}
Musical interaction with AI is often organised as a response loop: a human
performs, the system interprets that action, and the system answers,
accompanies, or schedules a musical event. Unconducted vocal ensembles pose a
different problem. Singers act simultaneously and continuously affect one
another; neither timing nor pitch is fixed by a conductor, metronome,
accompaniment, score, or tuning source. Collective organisation emerges from
many-to-many reciprocal adjustment. This paper frames such ensembles as
coupled dynamic systems and proposes a research architecture for vocal agents
that enter, rather than merely track, their collective states. Some target
repertoires are metrical, while others exhibit non-isochronous temporal
contours that cannot be reduced to a beat grid; we treat the latter as a hard
case for a general framework. The architecture connects multichannel capture
in the field to dialect- and singing-aware representation, collective-state
inference, vocal generation, and in-situ evaluation. The resulting agenda asks
not only whether an artificial singer can synchronise, but how its presence
reorganises human coordination, leadership, style, and musical transmission.
\end{abstract}



\setlength{\emergencystretch}{2em}
\raggedbottom
\maketitle

\section{Introduction}

Interactive music systems have established several powerful models of machine
musicianship. Voyager combined responsive and independent behaviour in a
nonhierarchical improvising orchestra, while the Continuator learned a
musician's style and generated continuations \cite{lewis2000,pachet2003}. More
recent systems include IRCAM's multi-agent Somax2 and the Metacreation Lab's
MACAT and MACataRT \cite{assayag2022,lee2024}. These systems differ
substantially, and surveys of musical agents show that no single interaction
model describes the field \cite{tatar2019}. Yet a recurring architecture
alternates machine listening with generation: a human action is interpreted
and the system responds, accompanies, or produces the next event. Even where
the machine has autonomy, influence is commonly organised around one human
input stream or a sequence of turns.

Unconducted vocal ensemble singing presents another interactional topology.
Several singers produce sound at once; each hears and affects the others while
predicting phrase progress, timing an entry or release, stabilising an
interval, shaping a vowel, and deciding whether to follow or pull. There may be
no conductor, accompaniment, metronome, fixed tuning source, written score, or
permanently privileged leader. Timing and pitch are not supplied to the group
from outside. They emerge through the group's activity.

This makes vocal ensembles a concentrated model system for collective
human--AI interaction. Their coordination is continuous rather than
turn-based, many-to-many rather than dyadic, and relational rather than
reference-based. A singer's slight anticipation may draw the group forward, be
absorbed without consequence, create a temporary split, or be reinterpreted as
a new phrase trajectory. The relevant computational object is therefore not
only a tempo estimate, score position, or aggregate synchrony value. It is the
changing structure of mutual influence through which a collective state is
produced.

We propose a research framework for an artificial singer that becomes another
node in this system. The near-term engineering path includes alignment, score
following, generation, and low-latency audio; the scientific aim is broader:
to model how each participant pulls and is pulled by the others, and to observe
how the network reorganises when an AI voice enters it. The paper develops this
dynamic formulation, treats non-isochrony as a demanding case, and sets out a
staged architecture grounded in VocalLanes data.

This paper presents a research framework, not a completed collective-state
model. VocalLanes currently records and aligns singer-dominant phone
recordings; symbolic representation is under development, while influence
modelling, agent policy, and comparative evaluation remain proposed. The
current observations come mainly from one ensemble performing traditional
Ukrainian village songs and motivate hypotheses for testing with further
groups. Our contribution is to define reciprocal coordination without an
external reference as a distinct problem and set out an empirically testable
route to an artificial singer within the ensemble's dynamics.

\section{Collective Dynamics without an External Reference}

In reference-based coordination, error can be measured against a beat grid,
score, backing track, conductor gesture, or designated leader. Such
representations remain useful when the reference is musically real. In an
unconducted ensemble, however, even a recognisable pulse is maintained
endogenously. Likewise, singers may stabilise intervals and sonorities relative
to one another while the ensemble's absolute pitch drifts. Relational coherence
can therefore remain high while every participant departs from an external
clock or tuning standard. A minimal formulation is:
\begin{equation}
  x_i(t+1) = f_i\bigl(x_i(t), x_{-i}(t-d), A_i(t), q(t), c\bigr).
\end{equation}
Here \(x_i(t)\) denotes singer \(i\)'s evolving state, including phrase-position
hypotheses, local duration and pitch expectations, uncertainty, and recent
observations of the other parts. The other-singer term denotes observations
received under perceptual and system delay \(d\); the incoming row of \(A(t)\)
represents time-varying influence weights from other participants to singer
\(i\); \(q(t)\) is a learned structural representation of the current song and
phrase; and \(c\) contains singer-, genre-, language-, and tradition-specific
constraints. No term is a privileged global clock. A collective state is
expressed jointly through the evolving influence network \(A(t)\), common and
singer-specific drift, uncertainty, clustering, and structural position.

This formulation separates phenomena that an average tempo would conflate. A
local change may be shared drift, an individual deviation, coordinated phrase
expansion, or recovery after uncertainty. Directional influence may change
from one phrase to the next. The state can therefore represent convergence,
distributed negotiation, and temporary divergence without assuming that these
form a universal taxonomy. Their definitions must remain open to the
categories and judgments of the singers who practice each tradition. The
closest precedent is the notion of an interpersonal synergy: a higher-order
system in which participants are coupled and compensate for one another rather
than executing independent plans \cite{riley2011}.

A useful representation must distinguish structural, individual, relational,
and collective state. Structural state describes where the performance is
within a learned song and what durations or variants are plausible there.
Individual state describes each singer's timing, pitch, and uncertainty.
Relational state describes pairwise lag, correction, and influence. Collective
state describes higher-order organisation in those relations, such as common
drift or distributed leadership. It is not an average of individual states; it
concerns the topology and evolution of their dependencies.

These levels unfold at different timescales. Consonant displacement or vowel
matching may occur over milliseconds, mutual adjustment over notes or
syllables, and phrase negotiation over several seconds. Repeated verses may
reveal slower adaptation. Hierarchical latent-state models are attractive
because they connect local observations to phrase-level organisation while
preserving uncertainty.

Signal-derived directionality is not causal ground truth: the same lag may
reflect shared song knowledge or response to a third singer. Inference must
therefore combine acoustics, musical structure, repeated takes, perturbation,
and singers' accounts. Alignment measures are useful, but successful
coordination remains repertoire- and practice-dependent \cite{keller2014}.

\section{Non-isochrony as a Hard Case}

In much Western art and popular music, rhythm is understood against an
isochronous metrical grid: a measure such as 4/4 is constructed from equal
underlying units. Musicians may deviate from their exact placement for
expression, but the equal units remain implied by notation and by performers'
metrical representation.

Across musical cultures, metrical and non-isochronous repertoires coexist.
Ukrainian and other Eastern European village singing traditions make the
contrast especially clear: lyric and dance genres may be metrical, while some
ritual genres organise successive syllables and phrase segments through stable
but unequal durations. Local time expands and contracts in a recurring
verse-level contour shaped by lyrics, prosody, ritual function, and style.
Performing such a contour against a steady beat can destroy rather than improve
coordination.

Non-isochrony exposes a hidden assumption in many accounts of synchrony: that
coordination is convergence around periodic time. In a non-isochronous phrase,
``ahead'' and ``behind'' are undefined unless the model also knows where the
ensemble is within a learned structural contour. It must distinguish
structural duration, global rate and drift, participant-specific timing, and
directional influence.

We therefore compare non-isochronous ritual songs with more metrical lyric
songs. Periodicity should be learned as a property of repertoire, not imposed
as a prerequisite of coordination. Hierarchical Semi-Markov Models or
Hierarchical Markov Renewal Processes can represent verses as sequences of
explicit duration states, while slower latent variables represent shared rate
and drift. This yields a probabilistic ``verse-shape'' rather than a beat grid
and extends accounts of temporal attending and entrainment
\cite{clayton2005,large1999}. IDyOM is a foundational probabilistic model of
musical expectation \cite{pearce2012}, including applications to auditory
boundary perception \cite{pearce2010}. IDyOT develops the broader
information-dynamics framework for online timing \cite{forth2016}.

Melodic variation, including improvised melody, is common in Ukrainian and
other Eastern European village singing traditions. Singers may vary melodic
lines and ornaments between verses while remaining recognisably within the
song and its style. This is not error around a canonical score. Pitch
coordination is likewise relational: singers negotiate intervals and
sonorities while absolute pitch may drift. The agent must therefore learn
melodic constraints, tradition-specific sonorities, and singer-specific
intonation and timbre without presuming equal temperament or a fixed reference
frequency.

\begin{figure}[t]
  \centering
  \includegraphics[width=0.673\columnwidth]{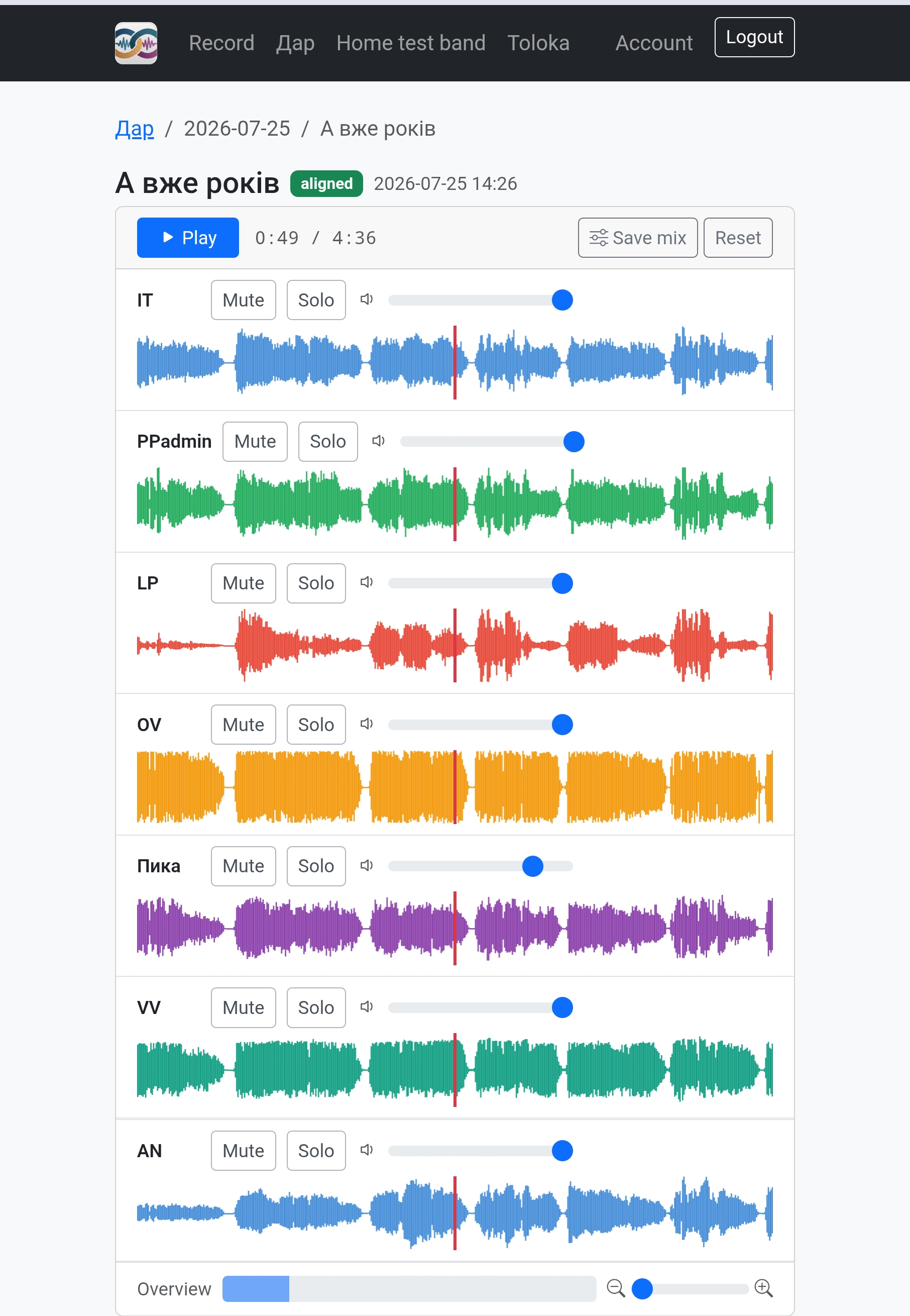}
  \caption{VocalLanes aligns phone recordings for mixed or foreground
    listening.}
  \Description{A mobile VocalLanes interface showing an aligned rehearsal take
    and seven colour-coded singer-dominant waveforms. Each track has mute, solo,
    and level controls, allowing the ensemble mix or an individual singer to be
    foregrounded during playback.}
  \label{fig:vocallanes}
\end{figure}

\section{VocalLanes: Rehearsal Tool and Data Infrastructure}

Collective-state models require synchronised observations of singers
interacting naturally. Clean stems are difficult to obtain because singers
must hear one another; a mixed recording, meanwhile, quickly masks individual
voices. Public multitrack collections are small or structurally mismatched:
the Choral Singing Dataset contains three SATB pieces and Dagstuhl ChoirSet two
pieces plus exercises \cite{cuesta2018,rosenzweig2020}; ESMUC and Cantoria add
fourteen songs \cite{cuesta2022}. JaCappella is larger, but its parts were
recorded separately \cite{nakamura2023}.

Field archives come closer. The Georgian Vocal Music corpus contains more than
200 performances, but access to its multitrack material is mediated
\cite{scherbaum2019}. The Polyphony Project documents thousands of Ukrainian
field performances, yet does not release reusable stems \cite{apjok2024}. No
existing corpus combines natural interaction, accessible singer-dominant
channels, repeated takes, lyrics, and coordination metadata.

VocalLanes addresses this through a rehearsal tool. Each singer records the
same performance on their own phone. Pairwise offset detection aligns most
tracks; DTW \cite{ewert2009} is reserved for unreliable cases. Recordings are
replayed as a mix or with any singer foregrounded, allowing singers to inspect
their part in context. The corpus contains 40 aligned takes from ten rehearsals
(about 2.3 hours): 39 from one ensemble in formations of three to seven
singers, and one four-singer take from a second ensemble. For the past two
months, the main ensemble has used VocalLanes routinely, with aligned playback
forming part of rehearsal practice rather than a separate data-collection
session. The prototype is not publicly available.

Pink-noise tests with known offsets all met the 40\,ms target. Reapplying the
detector to two aligned four-minute phone takes found residual offsets of
10--16\,ms and no drift; without an independent reference, this is an internal
consistency check rather than an accuracy evaluation. Over several months,
most outputs have supported comparative listening; a minority appear to
approach the roughly 30\,ms lag that becomes noticeable in rehearsal, although
this remains unmeasured.

The method adapts a field-recording practice used by Eastern European
ethnomusicologists: channels are singer-dominant but retain bleed, and repeated
takes arise organically. Consent and participant-controlled access are
implemented. The system is deliberately audio-first: lyrics and symbolic
structure complement the acoustic stream, while optional visual sensing would
reduce field compatibility.

\section{Barriers}

The scarcity and mismatch of data described above are not peripheral
inconveniences. Models must learn from few examples while handling bleed that
is intrinsic to live ensemble singing. Neither assuming clean stems nor
scaling a generic speech or music model resolves the target task.

Phonetics also behaves differently in singing. Vowels are prolonged and
ornamented; consonants may be displaced; melismas, breaths, and altered
transitions break ordinary speech-duration assumptions. Small languages and
dialects are poorly represented in training data. Many Voices demonstrates
systematic cross-cultural acoustic differences between song and speech
\cite{ozaki2024}, while singing-specific alignment reports greater
phoneme-boundary imprecision than for speech \cite{teytaut2021}. Lyrics
alignment is therefore an open modelling problem, not a solved preprocessing
step.

Singing also lacks the sharp, agreed onsets available in many instrumental
signals. A consonant, vowel nucleus, stable pitch, and perceived note beginning
may occur at different times; VocalNotes shows that even expert boundary
judgments vary across musical cultures \cite{proutskova2025}. The ambiguity
affects both alignment and score following. The original Antescofo formulation
targets scored instrumental performance \cite{cont2008}. Applying its
framework to voice required a bespoke HSMM extension combining melody and
lyrics, demonstrated on two solo professional singers \cite{gong2015}.

\begin{figure}[t]
  \centering
  \includegraphics[width=\columnwidth]{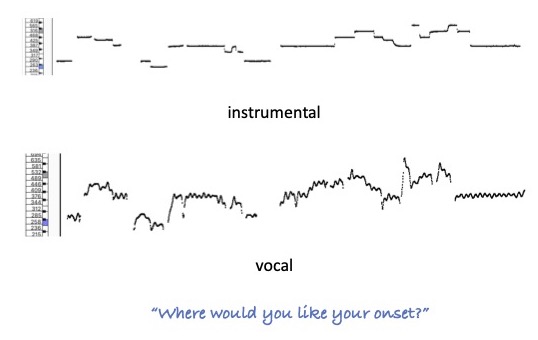}
  \caption{Instrumental note events (top) and a sung \(f_0\) contour (bottom)
    illustrate why onset location is ambiguous in singing
    \cite{proutskova2025}.}
  \Description{Two pitch displays are compared. The instrumental example has
    discrete notes with sharp beginnings, while the vocal fundamental-frequency
    contour is continuous, ornamented, and lacks unambiguous onset points.}
  \label{fig:onsets}
\end{figure}

Oral transmission creates a different representational barrier. Lyrics,
recurring verses, melodic formulas, and genre constraints may exist without a
canonical score. Symbolic representations must be inferred from repeated
performances rather than treated as prior ground truth. Note boundaries and
pitch categories must also remain culturally situated
\cite{proutskova2025}.

Finally, locating each singer does not reveal the collective state: influence
is latent, changes over time, and cannot be read from temporal precedence
alone. Real-time constraints compound the problem. Low-latency vocal analysis
is feasible \cite{ardaillon2019}, but real-time singing voice synthesis is not
yet available at the required quality and latency. These barriers define the
stages of the architecture rather than forming a separate list of technical
deficits.

\section{Proposed Method}

\begin{table}[t]
  \caption{Proposed method and implementation status.}
  \label{tab:method}
  \small
  \begin{tabularx}{\columnwidth}{@{}>{\raggedright\arraybackslash}p{0.29\columnwidth}>{\raggedright\arraybackslash}X@{}}
    \toprule
    Stage & Purpose \\
    \midrule
    Capture\newline\textit{implemented} &
      Singer-dominant phone recordings with bleed and automated alignment. \\
    \cmidrule(lr){1-2}
    Representation\newline\textit{in progress} &
      Dialect-aware lyrics, song structure, temporal contours, and melodic
      variation. \\
    \cmidrule(lr){1-2}
    State inference\newline\textit{proposed} &
      Singer position, uncertainty, shared drift, and changing influence. \\
    \cmidrule(lr){1-2}
    Agent/evaluation\newline\textit{proposed} &
      Online timing, vocal rendering, and reciprocal effects in real
      rehearsals. \\
    \bottomrule
  \end{tabularx}
\end{table}

The in-progress representation stage aims to infer symbolic and linguistic
structure from recordings, lyrics, and repeated performances. Singing-aware
vowel nuclei and dialectal grapheme-to-phoneme mapping will provide label
sequences, while Connectionist Temporal Classification (CTC) or
posteriorgram-DTW will estimate phoneme and syllable timing
\cite{teytaut2021,teytaut2023}. Comparisons across takes will recover verse
structure, temporal contours, and melodic formulas while separating singer
tendencies from genre rules.

These representations would support score following and collective-state
inference. The proposed system will combine \(f_0\), energy, and alignment
evidence to maintain position probabilities under bleed, while streaming
inference will separate shared from singer-specific drift. Influence from
singer \(j\) to singer \(i\) in \(A(t)\) will be supported when \(j\)'s recent
state improves prediction of \(i\)'s next event beyond \(q(t)\), shared drift,
and \(i\)'s history. This will represent predictive, not causal, influence.

The agent will condition melodic variation and vocal style on learned song and
singer representations. Until singing voice synthesis approaches 30\,ms
latency, the HSMM will generate timing online and drive a pre-generated vocal
line through time-stretching. Its influence will be graded by uncertainty: it
may join a shared phrase expansion, reduce its pull while leadership is
unresolved, or temporarily follow. Each action will affect the singers, whose
responses will enter the next inference step. Musical actions will be
co-designed with singers.

Prospective evaluation will compare repeated trials with human singers alone,
a reference-following voice, and a state-conditioned agent across metrical and
non-isochronous repertoire. Technical tests will ask whether cross-singer
histories improve held-out prediction beyond individual history, \(q(t)\), and
shared drift. Rehearsal analysis will compare inferred influence with singers'
accounts of control; VocalLanes will support post-take listening and
discussion.

\section{Safe and Responsible Innovation Statement}

Voice recordings are personally identifiable and can enable unwanted imitation
or cloning. VocalLanes already implements informed consent and
participant-controlled access. Scaling to further ensembles, and any use for
model training, synthesis, or release, requires explicit and separable
permissions. Models must not universalise one tradition's norms. Singers and
experts should shape annotations and agent behaviour, especially where war,
displacement, or cultural loss are involved. The purpose is to extend human
agency.

\section*{Generative AI Use Disclosure}

ChatGPT and OpenAI Codex supported drafting, editing, literature discovery, and
combining original screenshots into Figure~\ref{fig:vocallanes}. The author
reviewed all output and remains
responsible.

\bibliographystyle{ACM-Reference-Format}
\bibliography{references}

\end{document}